\documentclass[12pt]{article}

\begin{document}
\def\beq{\begin{eqnarray}}
\def\eeq{\end{eqnarray}}
\newcommand{\nn}{\nonumber}
\newcommand{\reals}{\mbox{${\rm I\!R }$}}
\newcommand{\nats}{\mbox{${\rm I\!N }$}}
\newcommand{\intgs}{\mbox{${\rm Z\!\!Z }$}}
\newcommand{\ujk}{u_{j,k}(1)}
\newcommand{\ujik}{u_{j,ik}(1)}
\newcommand{\uek}{u_{1,k}(1)}
\newcommand{\uzk}{u_{2,k}(1)}
\newcommand{\ueik}{u_{1,ik}(1)}
\newcommand{\uzik}{u_{2,ik}(1)}
\newcommand{\tk}{\frac{d}{dk}}
\newcommand{\pk}{\frac{\partial}{\partial k}}
\newcommand{\uja}{u_{j,k} (0)}
\newcommand{\ujb}{u_{j,k} (1)}
\newcommand{\vja}{v_{j,k} (0)}
\newcommand{\vjb}{v_{j,k} (1)}
\newcommand{\ujap}{u_{j,k}' (0)}
\newcommand{\ujbp}{u_{j,k}' (1)}
\newcommand{\ujx}{u_{j,k} (x)}
\newcommand{\vjx}{v_{j,k} (x)}
\newcommand{\ujxp}{u_{j,k}' (x)}
\newcommand{\vjxp}{v_{j,k}' (x)}
\newcommand{\ujxe}{u_{j,k}^{(1)} (x)}
\newcommand{\ujxz}{u_{j,k}^{(2)} (x)}
\newcommand{\ujxep}{u_{j,k}^{(1)\prime} (x)}
\newcommand{\ujxzp}{u_{j,k}^{(2)\prime} (x)}
\newcommand{\vjxe}{v_{j,k}^{(1)} (x)}
\newcommand{\vjxz}{v_{j,k}^{(2)} (x)}
\newcommand{\vjxep}{v_{j,k}^{(1)\prime} (x)}
\newcommand{\vjxzp}{v_{j,k}^{(2)\prime} (x)}
\newcommand{\drel}{\det (M+N H_{1,k} (1))}
\newcommand{\sca}{k^2 \langle u_{1,0} | u_{1,k} \rangle }
\newcommand{\scaneu}{k^2 \langle u_{1,0} | v_{1,k} \rangle }

\title{Functional determinants by contour integration methods}
\author{Klaus Kirsten\\
$ $\\
Max Planck Institute for Mathematics in the Sciences,\\
Inselstrasse 22-26, 04103 Leipzig, Germany\\
and\\
Mathematics Department, Baylor University,\\
Waco TX 76798, USA\\
$ $\\
Alan J. McKane\\
$ $\\
Department of Theoretical Physics,\\
University of Manchester, Manchester M13 9PL,\\
 England}
\date{\today}
\maketitle
\begin{abstract}

We present a simple and accessible method which uses contour
integration methods to derive formulae for functional
determinants. To make the presentation as clear as possible, the
general idea is first illustrated on the simplest case: a second
order differential operator with Dirichlet boundary conditions.
The method is applicable to more general situations, and we
discuss the way in which the formalism has to be developed to
cover these cases. In particular, we also show that simple and
elegant formulae exist for the physically important case of
determinants where zero modes exist, but have been excluded.

\end{abstract}

\newpage

\section{Introduction}
\label{intro}

The need to calculate functional determinants, usually of second order
differential operators, arises in many different fields. For example,
the leading order contribution to path-integrals often involves the
integration of the exponential of a quadratic form, which can be carried
out exactly, and yields a functional determinant~\cite{fey65,sch81}.
The explicit formulae for one-dimensional functional determinants of this
kind are not simply of interest because of their utility, but also because
rather general results may be obtained. For instance, the functional
determinant of an operator $L$ can often be given explicitly in terms of
the solutions of the homogeneous equation $Ly=0$. The elegance and generality
of the results that can be obtained has attracted mathematicians, especially
in the last decade or so. However, many researchers who are simply searching
for an explicit form for a particular differential operator may find these
treatments rather abstract, and, in addition, some will be interested in the
value of these determinants with the zero modes extracted. These zero modes
will usually exist whenever a quadratic form in a path-integral is the result
of performing a linearisation about a solution with non-trivial spatial or
temporal dependence. By Goldstone's theorem, this symmetry breaking will give
rise to zero modes, which must be excluded from the quadratic form if this is
to be a consistent leading order approximation~\cite{raj82}. Thus the
functional determinant obtained by evaluating the path-integral is as before,
but with these modes absent.

This paper is designed to give a simple, and easily accessible,
derivation of the main formulae for functional determinants, and
also to show how these results may be modified to give analogous
expressions for determinants with the zero mode (or modes)
extracted. We will avoid unnecessary mathematical abstraction,
concentrate on the most straightforward results and give examples
to illustrate the methods; a forthcoming paper will discuss some
of the technical points in more detail~\cite{kir03}. Specifically,
we will use only contour integration methods and elementary facts
concerning the solutions of differential equations and concentrate
on second order differential operators. An earlier study of
determinants with the zero modes extracted~\cite{mck95} obtained
results by invoking a regularisation procedure in which the zero
modes became non-zero, extracting them, and then removing the
regularisation. The problem with this approach is that it is not
clear that the result is independent of the regularisation. Within
the contour integration adopted here, the zero modes are not
included from the beginning, and there is no need to introduce any
regularisation procedure for them.

An early review of functional integrals and their evaluation~\cite{gel60}
gives a prescription for finding the functional determinant of a second order
differential operator with no first order derivative and with Dirichlet
boundary conditions. The method consists of discretising the functional
integral, explicitly carrying out the resulting multidimensional Gaussian
integral, and showing that, on return to the continuum, the determinant
satisfies the homogeneous equation with given boundary conditions. This
method is also clearly presented in Schulman's book on
path-integrals~\cite{sch81}. During the 1970's quantum field theory
in general, and functional integral techniques in particular, became very
popular and many studies of non-trivial field configurations such as solitons
and instantons appeared. In some cases~\cite{bre77} the calculation of the
determinants involved in these studies were carried out by explicitly
determining the eigenvalues of the Sturm-Liouville problem and calculating
their product. This had the advantage that the omission of the zero mode
required in these cases could be carried out explicitly. On the other hand,
Coleman, in an appendix to his 1977 lectures on instantons at
Erice~\cite{col85}, gives the outline of a rather straightforward
demonstration of the result for Dirichlet boundary conditions mentioned above.
The method, which depends on the result being known a priori, consists of
shifting the eigenvalues by a parameter $z$, and arguing that both sides of
the result to be proved are analytic in $z$ and have the same zeros. Standard
results from the theory of complex variables then allows the result to be
proved.

The first published studies devoted to obtaining formulae for functional
determinants of this kind were by Levit and Smilansky~\cite{lev77} and
Dreyfus and Dym~\cite{dre78}. They extended the known results to more general
kinds of operators, for example those of higher order, but continued to assume
Dirichlet boundary conditions. While Dreyfus and Dym used a method of proof
similar to Coleman's, Levit and Smilansky proved the result by showing that
the logarithmic derivatives of both sides of the required result were equal.
This was also the approach adopted by Forman~\cite{for87}, who extended the
known results to situations with more general boundary conditions. These
results were rederived by yet another method a few years later~\cite{for92}.
Burghelea {\em et al} investigated these and closely related questions using
more abstract language~\cite{bur91,bur93} (see also~\cite{car02});
Ref \cite{bur95} contains an accessible discussion of some of their results.
In particular unknown prefactors are determined in these articles. Other
authors~\cite{sar87}-\cite{bar94} were concerned with defining generalised
forms of determinants (the ``p-determinant'' or the ``$\Delta$-determinant'')
which act as regularised versions of the conventional determinant, and may be
useful in cases where the usual determinant does not exist. Recent work has
extended some of this rigorous work to regular singular
potentials~\cite{les98a} and to more general kinds of boundary
conditions~\cite{les98b}. Also explicit formulae that relate quotients of
determinants with the corresponding Green function have been
given~\cite{fal99}.

In contrast to the many different techniques devised to prove
these rather mathematically attractive results, the task of
obtaining analogs for functional determinants with the zero mode
extracted has been neglected. Apart from the paper by McKane and
Tarlie already mentioned~\cite{mck95}, the only published articles
to address this point have been by Kleinert and
Chervyakov~\cite{kle98,kle99}. They base their approach on the
Wronski method for constructing Green functions, but they do not
discuss cases where there is more than one variable, and again
introduce an explicit regularisation as an intermediate step in
the calculation. In this article we will use the contour
integration methods developed in~\cite{bor96a,bor96b} to avoid
introducing ad hoc regularisation procedures for the zero modes.
For a recent review of these methods see Kirsten~\cite{KKbook}.
This will allow us to easily obtain the formulae for functional
determinants we seek in terms of solutions of the associated
homogeneous equation and the boundary data of the problem. The
results generalise known answers where the absence of zero modes
was assumed.

The outline of the paper is as follows. In Section \ref{simple} we describe
the method in the most straightforward case: obtaining the functional
determinant of a second order differential operator with Dirichlet boundary
conditions. Having illustrated the method, in Section \ref{zeromode} we
discuss how the formalism has to be modified to give results when any zero
mode which exists has been omitted. In sections \ref{general} and
\ref{general_zeromode} we extend the treatment of sections \ref{simple} and
\ref{zeromode} respectively, to general boundary conditions. The extension
to systems of differential equations will be briefly discussed in section
\ref{systems} but, for simplicity, we will limit ourselves to an outline and
give results which for the most part apply to a specific example. We conclude
in section \ref{conclusion}. Two appendices contain technical results
pertaining to the development of some aspects of the formalism described in
sections \ref{general_zeromode} and \ref{systems}.

\bigskip

\section{Dirichlet boundary conditions}
\label{simple}

We begin by considering the simplest class of differential
operators: \beq L_j = -\frac{d^2}{dx^2} + R_j  (x)\label{start1}
\eeq on the interval $I=[0,1]$. Since the determinant of any such
operator will diverge, it is usual to calculate ratios of
determinants both of which involve operators having similar
structure, for instance, defined on the same interval and having
the same leading order term. The index $j=1,2$ labels these two
determinants. We will adopt the convention that the $j=1$ operator
is the one for which we wish to calculate the determinant and the
other, $j=2$ one, will be some standard reference operator, in
terms of which our result can be given. As long as the interval in
the problem of interest is finite, we can shift and scale $x$ so
that $x \in [0,1]$. The operators in this section will have no
first derivative term and no non-trivial function of $x$
multiplying the second derivative term. We will factor out any
constant terms from the $L_j$ so that the coefficients of the
second derivatives is $-1$. The functions $R_j (x)$ will be
assumed to be continuous and all the eigenvalues will be assumed
to be positive.

In applications which involve the evaluation of functional integrals, ratios
of determinants frequently arise, as do Dirichlet boundary conditions. So it
is natural that the first results which were obtained should have been for
this case. As we have already mentioned in Section \ref{intro}, the result
involves the solutions of the associated homogeneous equation. In fact it
is given in terms of the solution which satisfies given initial conditions:
\beq
\frac{\det L_1}{\det L_2} = \frac{y_1 (1)}{y_2(1)} ,\label{1}
\eeq
where $y_j (x)$ is the unique solution of
\beq
L_j y_j (x) = 0, \quad y_j (0) = 0, \quad y_j ' (0) =1 . \nn
\eeq
The main purpose of this section is to rederive this result by a direct
contour integration method.

To do this let $u_{j,k}(x)$ be the unique solution of 
\beq 
(L_j - k^2) u_{j,k} (x) =0  
\nn 
\eeq 
satisfying 
\beq 
u_{j,k} (0) = 0, \quad u_{j,k} ' (0) = 1 , 
\label{iniDir} 
\eeq 
$k$ being a complex parameter. The eigenvalues are then fixed by imposing 
\beq 
\ujk =0. 
\label{eigencond_Dir} 
\eeq 
The second condition in (\ref{iniDir}) is simply a choice of normalisation 
for the eigenfunction $u_{j,k} (x)$; we will indicate at the end of this
section what happens if we make a more general choice. In what follows we
will assume that $u_{j,k} (1)$ is analytic in $k$ for $\Re k \geq 0$. The 
definition of this eigenvalue problem allows the zeta function of $L_j$ to 
be written as 
\beq 
\zeta_{L_j} (s) = \frac 1 {2\pi i} \int_\gamma dk k^{-2s}
 \tk \ln \ujk . 
\nn
\eeq 
The contour $\gamma$ is counterclockwise and encloses all
eigenvalues on the positive (by assumption) real axis. As given,
the representation is valid for $\Re s > 1/2$. Clearly
\beq
\zeta_{L_1} (s) -\zeta_{L_2} (s)
       = \frac 1 {2\pi i} \int_\gamma dk k^{-2s}
 \tk \ln \frac{\uek}{\uzk} .
\label{2}
\eeq

The next step is to deform the contour to the imaginary axis. In the case
being considered here there are no negative or zero modes, this means
$u_{j,0} (1) \neq 0$, and so no contribution arises from the origin.
The leading $|k|\to \infty$ behaviour is independent of $R_j (x)$ and is
governed by \cite{cou53}
\beq
u_{j,k} (x) \sim \sin (kx) \left(
   1 + {\cal O} (k^{-1}) \right) , \label{CHas}
\eeq $j=1,2$, so that for $|k| \to \infty$ we have \beq \tk \ln
\frac {u_{1,k} (1)} {u_{2,k} (1) } = {\cal O } (k^{-2}) . \nn \eeq
With $\epsilon >  0$ a small real number introduced to indicate
that the contour approaches the imaginary axis from the right,
this allows us to write \beq \zeta _{L_1} (s) -\zeta_{L_2} (s) =
\frac 1 {2\pi i} \int _\infty ^{-\infty} dk (ik+\epsilon) ^{-2s}
\tk \ln \frac{\ueik}{\uzik}, \label{3} \eeq which is valid for
$\Re s > -1/2$.

The next relevant observation is that $u_{j,ik}(x)$ and
$u_{j,-ik}(x)$ satisfy the same differential equation, and since
they are the unique solution one has $u_{j,ik}(x) = u_{j,-ik}(x)$.
The integral in (\ref{3}) may be broken up into two contributions,
$k>0$ and $k<0$. The two contributions combine to give \beq \zeta
_{L_1} (s) -\zeta_{L_2} (s) =\frac{\sin (\pi s)}\pi \int_0^\infty
dk k^{-2s} \tk
 \ln \frac{\ueik}{\uzik} ,\nn
\eeq which is valid for $-1/2 < \Re s < 1/2 $. The upper
restriction comes from the lower integration bound $k=0$, whereas
the lower restriction is from the upper integration limit as
discussed.

From here it is immediate that \beq \zeta ' _{L_1} (0) - \zeta '
_{L_2} (0) = \int_0^\infty dk \tk
 \ln \frac{\ueik}{\uzik} = -\ln \frac{u_{1,0} (1)}{u_{2,0} (1)}
 = -\ln \frac{y_1
(1)}{y_2 (1)} ,\nn \eeq
which reproduces eq.~(\ref{1}) using the
definition \beq \det (L_1 L_2^{-1}) = \exp{-(\zeta_{L_1 }'(0) -
\zeta ' _ {L_2} (0))}. \label{det_defn} \eeq

If we do not normalise $u_{j,k}' (0) =1$ but instead ask that
$u_{j,k}' (0) = c_j$, the above integral will give a contribution
at infinity: \beq \zeta ' _{L_1} (0) - \zeta ' _{L_2} (0) = \ln
\frac{u_{1,i\infty} (1)}{u_{2,i\infty} (1)} - \ln \frac{u_{1,0}
(1)}{u_{2,0} (1) } = -\ln\frac{c_2 u_{1,0}(1)}{c_1 u_{2,0}(1)} ,
\nn \eeq because now \beq \frac{u_{1,i\infty} (1)} {u_{2,i\infty}
(1) } = \frac{c_1}{c_2} .\nn \eeq This is consistent with the
answer expected: \beq \det (L_1 L_2 ^{-1}) = \frac{c_2 y_1 (1)}
{c_1 y_2 (1)} = \frac{y_1 (1) y_2 ' (0)} {y_1 ' (0) y_2 (1)}.
\label{normdet} \eeq The strength of the method is that, as
discussed later, it can be applied with very little modification
to more complicated operators and boundary conditions. However,
just as important, we can use it to determine the analogous result
to (\ref{normdet}) for the situation where a zero mode is present
and has been extracted.

\bigskip

\section{Omitting the zero mode}
\label{zeromode}

To explain how the procedure discussed in the last section can be modified
to deal with a situation in which a zero mode is present, but has been
omitted from the evaluation of the determinant, we stay with the simple
operator (\ref{start1}) and Dirichlet boundary conditions. However, as before,
we will see that the method generalises to more complicated situations.
In addition to (\ref{iniDir}) we now also suppose that $u_{1,0}(1)=0$, that
is, $L_{1}$ has an eigenfunction with zero eigenvalue, and all the other
eigenvalues are assumed to be positive. The integrand in (\ref{2}) will now
have a pole at $k=0$ and therefore the contour $\gamma$ can no longer be
deformed along the imaginary axis. Our approach will be to consider a modified
form of the integrand which has no pole, and which still gives us the
determinant of $L_{1}$, but now with the zero eigenvalue extracted.

We first need to determine the behaviour of $u_{1,k}(1)$ for small
$k$ in order to eliminate the pole. To do this, we note that
integrating the left-hand side of 
\beq 
\int^{1}_{0} dx\,u_{1,0}(x)^* L_{1} u_{1,k}(x) = k^{2} \int^{1}_{0}
dx\,u_{1,0}(x)^* u_{1,k}(x) , 
\nn 
\eeq by parts gives 
\beq 
\left[ u'_{1,0}(x)^* u_{1,k}(x) - u'_{1,k}(x)^* u_{1,0}(x)
\right]^{1}_{0} + \int^{1}_{0} dx\,u_{1,k}(x) (L_{1} u_{1,0}(x))^*
= k^{2} \langle u_{1,0} | u_{1,k} \rangle . 
\nn 
\eeq 
Here $*$ denotes complex conjugation and we have introduced the Hilbert
space product $\langle \ | \ \rangle$ on ${\cal L}^2 (I)$ by 
\beq 
\langle u | v \rangle = \int _0 ^1 dx \,\, u(x) ^* \,\, v(x) . 
\nn
\eeq
Although we will be mostly concerned with real functions in this paper, the 
generalisation to complex functions will be required in section \ref{systems}, 
and so it is useful to allow for this possibility in the development of the 
formalism.

Using the boundary conditions this gives
\beq
u_{1,k}(1) = \frac{k^{2} \langle u_{1,0} | u_{1,k}
\rangle}{u'_{1,0}(1)^*} \equiv  - k^{2} f_{1,k} .
\label{4}
\eeq
Since $f_{1,k}$ is finite and non-zero as $k \rightarrow 0$, we
have the desired behaviour of $u_{1,k}(1)$. The minus sign has been
included in the definition of $f_{1,k}$ for later convenience. It is important
to note that (\ref{4}) is true for any $k$ --- no small $k$ assumption was
made to derive it.

We can now modify the discussion of section \ref{simple} to cover the
case when a zero mode is present, by using the following two observations:
\begin{itemize}
\item[(i)] The function $f_{1,k}$, defined by (\ref{4}), vanishes at all
values of $k$ for which $k^2$ is a positive eigenvalue, as $u_{1,k} (1)$
does. Only in the case of the zero eigenvalue do we have the situation where
$f_{1,0} \neq 0$, but $u_{1,0} (1)=0$. Therefore if we use $f_{1,k}$ instead
of $u_{1,k} (1)$ in (\ref{2}), and choose the contour so as to surround the
positive eigenvalues, the formalism developed in section \ref{simple} can be
used to obtain the required result.

\item[(ii)] The notion that $u_{1,k} (1)$ may be straightforwardly substituted
by $f_{1,k}$ has to be amended slightly, since these functions have different
behaviours as $k \to \infty$. This is because the method of proof we used in
the last section relies on $u_{1,ik} (1)$ and $u_{2,ik} (1)$ having the same
behaviour in this limit. This difference in behaviour is accounted for if we
replace $u_{1,k} (1)$ not by $f_{1,k}$, but by $(1-k^{2})f_{1,k}$, since then
$u_{1,ik} (1) \sim k^{2}f_{1,ik}$ as $k \to \infty$, while $(1-k^{2})f_{1,k}$
remains non-zero at $k=0$.
\end{itemize}

These remarks lead us to consider the contour integral
\beq
\frac 1 {2\pi i} \int_\gamma dk k^{-2s}\tk \ln (1-k^{2})f_{1,k} ,
\label{contour}
\eeq
where the contour surrounds all of the eigenvalues on the positive $k$ axis.
This integral equals
\beq
\zeta_{L_1} (s) + \frac 1 {2\pi i} \int_\gamma dk k^{-2s}
\tk \ln (1-k^{2}) ,
\label{contour_res}
\eeq
where it is understood that the zero mode has been omitted from the definition
of the zeta-function. The second term in (\ref{contour_res}) is equal to $1$
if the contour surrounds $k=1$, and zero if it does not. Our final result
does not depend on this choice so, for definiteness, we choose the contour to
surround this point.

The results (\ref{contour}) and (\ref{contour_res}) can now be
combined with the conventional result for the zeta function of
$L_{2}$ (the eigenvalues of $L_{2}$ are assumed to be all
positive). The contour can safely be deformed to the imaginary
axis and the contribution from infinity in the right-half plane
shown to vanish. Using $f_{1,ik} = f_{1,-ik}$ and proceeding as
before we find \beq \zeta _{L_1} (s) -\zeta_{L_2} (s) =\frac{\sin
(\pi s)}\pi \int_0^\infty dk k^{-2s} \tk \ln
\frac{(1+k^{2})f_{1,ik}}{\uzik} - 1. \label{6} \eeq The conditions
on $s$ for this representation to be valid are the same as those
given in section \ref{simple}, and therefore we may differentiate
with respect to $s$ and set $s=0$. If we impose the condition
$u_{j,k} ' (0)=1$, then the argument of the logarithm in (\ref{6})
equals $u_{1,i\infty} (1)/u_{2,i\infty} (1) = 1$ in the limit $k
\to \infty$. Therefore the contribution from the upper limit of
integration is zero. So only the contribution from the lower limit
remains and we obtain 
\beq 
\zeta ' _{L_1} (0) - \zeta ' _{L_2 } (0) = -\ln \frac{f_{1,0}}{u_{2,0} (1)} . 
\nn 
\eeq 
Using (\ref{det_defn}) and (\ref{4}) this gives 
\beq 
\frac{\det ' L_{1}}{\det L_{2}} = \frac{f_{1,0}}{u_{2,0} (1)} = 
- \frac{\langle y_{1} | y_{1} \rangle} {y'_{1}(1)^*  y_2 (1)} , 
\label{primed1}
\eeq 
where we have denoted the determinant of $L_1$ with the zero
mode extracted by $\det ' L_{1}$.

This result should be compared to (\ref{1}), rather than
(\ref{normdet}), since we have imposed the normalisation condition
$y_{j}'(0)=1$. If we had not imposed normalisation conditions, an
extra factor of $c_{1}/c_{2}$ would be present as in
(\ref{normdet}). We see that the modification required when a zero
mode is extracted gives a final result which is still very simple,
involving only the derivative of the zero mode at an end-point
(both end-points if the function is not normalised) and its norm.
It agrees with the result derived in Ref. \cite{mck95}, once
typographical errors in eqns. (2.7) and (2.14) of that paper have
been corrected. However, when making the comparison it should be
borne in mind that the operators used in that paper are minus
those used here, and so the extraction of a mode gives a result
which differs from ours by a sign. We can check that our result
has the correct sign by noting that the lowest eigenfunction
should have no nodes, so $y_{1} ' (0)$ and $y_{1} ' (1)$ should
have opposite signs. In particular, if we impose the condition
$y_{1} ' (0) = 1$, then $y_{1} ' (1) < 0$. Therefore,
(\ref{primed1}) should be positive, which is correct, since all
eigenvalues are positive.

In applications it may not even be necessary to calculate
$\langle y_{1} | y_{1} \rangle$. This is because, if the zero mode has been
extracted using the method of collective coordinates~\cite{raj82}, a Jacobian
appears due to the transformation to new coordinates which, to lowest order,
is proportional to the norm of the zero mode, and cancels it out. A little
care is required with this cancellation, since the zero mode appearing in the
Jacobian function will have a definite normalisation associated with it, which
it inherited from the classical solution (for instance, the zero mode will be
the spatial derivative of the classical solution if translational invariance
has been broken). Therefore an overall constant may have to be factored out
of the zero mode to get $y_{1}$, and similarly from the Jacobian to get
$\langle y_{1} | y_{1} \rangle$, but these are purely algebraic
procedures --- no integration will be required.

To illustrate the ideas presented so far in this paper, let us give an
example. A rich source of operators of the type (\ref{start1}) are
Hamiltonian operators in one-dimensional quantum mechanics. The functions
$R_{j}(x)$ will be the potentials; we will take as our example the linear
potential~\cite{lan77} $R_{1}(x) = x-x_{0}$ and normalise it with respect to
the case $R_{2}(x) = 0$. This potential has been chosen purely on grounds of
simplicity; it has not to our knowledge occurred in a calculation of
fluctuations about a nontrivial solution as described in the Introduction.
As a consequence, if we wish the lowest eigenvalue to be zero, the constant
$x_{0}$ has to be appropriately tuned.

The general solution of the equation $u_{1}'' = (x-x_{0}-k^{2})u_{1}$ which
satisfies $u_{1,k}(0)=0$ is, up to normalisation,
\beq
u_{1,k}(x) = {\rm Ai}(-z)\,{\rm Bi}(x-z) - {\rm Bi}(-z)\,{\rm Ai}(x-z) , \ \
z \equiv x_{0} + k^{2} ,
\label{airyk}
\eeq
where Ai and Bi are Airy functions~\cite{abr65}. Implementation of the second
boundary condition, $u_{1,k}(1)=0$, fixes the condition for $k^2$ to be an
eigenvalue. This condition takes the form of a transcendental equation
involving Airy functions, however a good approximation to the  allowed values
of $k^2$ may be obtained by using the asymptotic form for Airy functions, valid
when the argument of the function is large~\cite{abr65}. One finds
$z^{3/2} = (z-1)^{3/2} + 3n\pi/2,\ n=1,2,\ldots$, which implies that
$z \approx n^{2}\pi^{2} + 1/2$. Although this result has been formally derived
under the assumption that $z$ is large, it turns out to be an excellent
approximation even for quite small $n$: correct to 0.01\%. For the $n=1$
case, which is the least accurate, the true result is $10.3685\ldots$ and the
above approximation gives $\pi^{2} + 0.5 = 10.3696\ldots$. Note that for
large $n$ the eigenvalue condition becomes $z \approx n^{2}\pi^{2}$, which
is the result which would be found in the case $R_{2}(x)=0$. Thus it is at
least plausible that the ratio of the product of eigenvalues converges.
However, the whole point of the formalism we have been developing is to bypass
the determination of the eigenvalues, and simply work with the solutions of
the corresponding homogeneous equation.

The relevant solution of the homogeneous equation for this
particular example is $y_{1}(x)=\lim_{k \to 0} u_{1,k}(x)$, where
$u_{1,k}(x)$ is given by (\ref{airyk}), but subject to the
normalisation condition $u_{1,k}'(0)=1$, as in (\ref{iniDir}).
Thus
\beq y_{1}(x) &=& \frac{{\rm Ai}(-x_{0})\,{\rm Bi}(x-x_{0}) -
{\rm Bi}(-x_{0})\,{\rm Ai}(x-x_{0})}{{\rm Ai}(-x_{0})\,{\rm
Bi}'(-x_{0}) - {\rm Bi}(-x_{0})\,{\rm Ai}'(-x_{0})} \nn\\
&=&\pi \left[ {\rm Ai} (-x_0){\rm Bi} (x-x_0) - {\rm Bi} (-x_0) {\rm Ai}
(x-x_0) \right] ,
\label{airyx0}
\eeq
where we used the result for the
Wronskian determinant \cite{abr65}
\beq
W \left\{ {\rm Ai} (z) , {\rm Bi} (z) \right\} =
\frac 1 \pi .\nn\eeq Similarly, \beq y_{2}(x) = x .
\label{soln_2}
\eeq
Therefore using (\ref{2}),
\beq
\frac{\det L_1}{\det L_2} = \pi \left[ {\rm Ai} (-x_0) {\rm Bi} (1-x_0)
- {\rm Bi} (-x_0) {\rm Ai} (1-x_0) \right].
\label{det_airyx0}
\eeq
If $x_{0}=0$,
the result simplifies considerably~\cite{abr65}: \beq \frac{\det
L_1}{\det L_2} = \frac{\Gamma(1/3)}{2\,3^{1/6}} \left[ {\rm Bi}(1)
- \sqrt{3} {\rm Ai}(1) \right] = 1.085 . \label{det_airy0} \eeq

Now suppose that we artificially fix $x_0$ to be
$x^{*}_{0}=10.3685\ldots$ --- the smallest value of $z$ for which
$u_{1,k}(x)$ given by (\ref{airyk}) satisfies $u_{1,k}(1)=0$. Then
$k^{2}=0$ is an eigenvalue, $y_{1}(1)=0$ and the expression in
(\ref{det_airyx0}) is identically zero. To find the ratio of
functional determinants with the zero mode extracted from the
determinant in the numerator, we use (\ref{primed1}) and
(\ref{airyx0}) to find
\beq
\frac{\det ' L_{1}}{\det L_{2}} =
- \frac{ \langle y_{1} | y_{1} \rangle } {\pi ({\rm Ai} (-x_0^*)
{\rm Bi} ' (1-x_0^*) - {\rm Bi} (-x_0^*){\rm Ai} ' (1-x_0^*)
)}= 0.050666.
\label{det_airyx0star}
\eeq
Let us mention that $\langle y_1|y_1 \rangle$ can be expressed as a
combination of Airy functions and their derivatives. However, no
simplifications occur and we do not display the resulting answer.

\section{General boundary conditions}
\label{general}

In the last two sections we have illustrated how an expression for
the ratio of functional determinants can be derived in the simple
case of a single differential equation of the type (\ref{start1})
and Dirichlet boundary conditions. In the remainder of the paper
we will extend this treatment to more general situations. Those
parts of the proof involving zeta-functions and contour
integration will remain essentially unchanged; the novel aspects
arise from setting up the problem so that the method naturally
generalises. Since the main aim of this paper is to provide a
transparent discussion available to a wide readership, we will
limit our discussion to more general boundary discussions, leaving
the topic of more complicated second order operators to a future
paper~\cite{kir03}. We will, however, briefly discuss a system of
two coupled operators of type (\ref{start1}) in section
\ref{systems}, in order to provide some indication of how the
procedure generalises in that direction.

The boundary conditions on the functions on which $L_j$ operates
fall into two classes: separated and non-separated. Separated
boundary conditions are defined as those which do not mix
conditions on $\ujx$ and $\ujxp$ at the boundary $x=0$ with
conditions at the boundary $x=1$. Robin boundary conditions $A\uja
+B\ujap =0, C\ujb +D\ujbp =0$, where $A,B,C,D$ are given real
constants, are the generic case. Non-separated conditions can
involve $\ujx$ and its derivative at different boundaries in the
same equation. Periodic boundary conditions $\uja = \ujb, \ujap =
\ujbp$ are a common example. In previous sections, where Dirichlet
boundary conditions were imposed, initial conditions were
specified at $x=0$ which guaranteed a unique $\ujx$ on which to
impose boundary conditions. The imposition of more complicated
types of boundary conditions, and in particular those of the
non-separated type, does not permit us to proceed in the same way,
and we need to set up a more systematic approach.

The first modification consists of no longer working with the second order
differential equation
\beq
\frac{d^{2}u_{j,k}}{dx^2} = \left( R_{j}(x) - k^{2} \right) u_{j,k} ,
\label{second_order}
\eeq
but going over to the first order formalism where we define
$\vjx \equiv \ujxp$, and view the column vector
\beq
{\bf u}_{j,k}(x) = {\ujx \choose \vjx}
\label{first_order_ele}
\eeq
as the basic element of the theory. Then (\ref{second_order}) may be written as
\beq
\frac d {dx} {\ujx \choose \vjx} =
\left(
\begin{array}{cc}
 0 & 1\\
R_j -k^2 & 0
\end{array}
\right) { \ujx \choose \vjx} . 
\label{first_order} 
\eeq 
The general form for the boundary conditions can then be expressed in
terms of ${\bf u}_{j,k}(x)$, using the notation of Ref.
\cite{for87}, as 
\beq 
M  {\uja \choose \vja } + N { \ujb \choose \vjb} = {0 \choose 0} , 
\label{gbc} 
\eeq 
where $M$ and $N$ are $2\times 2$ matrices. For example, in the case of 
Dirichlet boundary conditions 
\beq 
M=\left(
\begin{array}{cc}
1 & 0 \\
0 & 0
\end{array}
\right) , \quad
N=\left(
\begin{array}{cc}
0 & 0 \\
1 & 0
\end{array}
\right) . \label{MN_Dir}
\eeq
Now let ${\bf u}_{j,k}^{(1)}(x)$ and ${\bf u}_{j,k}^{(2)}(x)$ be two
independent solutions of the differential equation (\ref{first_order}). We
will not impose the boundary condition (\ref{gbc}) on these solutions, instead
they will be uniquely specified by the imposition of initial conditions. They
may then be thought of as basis functions from which a general solution can be
constructed:
\beq
{\ujx \choose \vjx} &=& {\alpha \ujxe + \beta \ujxz \choose \alpha \vjxe
+ \beta \vjxz} \nn \\
&=& \left(
\begin{array}{cc}
u_{j,k}^{(1)} (x) & \ujxz \\
\vjxe & \vjxz
\end{array}
\right)
\left(
\begin{array}{c}
\alpha \\
\beta
\end{array}
\right) .
\label{lin_comb}
\eeq
In the special case when $k^2$ is an eigenvalue, the constants $\alpha$ and
$\beta$ can then be chosen so that ${\bf u}_{j,k}(x)$ satisfies the boundary
conditions (\ref{gbc}).

Now
\beq
H_{j,k} (x) =
\left(
\begin{array}{cc}
u_{j,k}^{(1)} (x) & \ujxz \\
\vjxe & \vjxz
\end{array}
\right)
\label{funda1}
\eeq
is a fundamental solution to (\ref{first_order}). Since
$\det H_{j,k}(x) \neq 0$,
\beq
{\alpha \choose \beta } = H_{j,k}^{-1} (0) {\uja \choose \vja} .
\nn
\eeq
So, in summary,
\beq
{\ujx \choose \vjx } = H_{j,k} (x) H_{j,k} ^{-1} (0)
  {\uja \choose \vja}
\nn
\eeq
is the unique solution of (\ref{first_order}) with initial value
$(\uja , \vja )$ at $x=0$. This allows us to rewrite the boundary condition
(\ref{gbc}) as
\beq
(M+N H_{j,k} (1) H_{j,k} ^{-1} (0) ) {\uja \choose \vja} =0 .
\label{gbc1}
\eeq
Nontrivial solutions only exist if
\beq
\det \left( M+N H_{j,k} (1) H_{j,k} ^{-1} (0) \right) =0 ,\label{nontri1}
\eeq
and this gives the condition for $k^2$ to be an eigenvalue.

Although we could in principle work with any set of
$\{ {\bf u}_{jk}^{(1)}(x),{\bf u}_{jk}^{(2)}(x) \}$, a particularly suitable
choice is
\beq
H_{j,k} (0) =
\left(
\begin{array}{cc}
u_{j,k}^{(1)} (0) & u_{j,k}^{(2)} (0) \\
v_{j,k}^{(1)} (0) & v_{j,k}^{(2)} (0)
\end{array}
\right)
=
\left(
\begin{array}{cc}
1 & 0 \\
0 & 1
\end{array}
\right) . \label{simini}
\eeq
This has a number of simplifying features. For instance, $\det H_{j,k}(x)=1$
for all $x$, $\alpha=\uja, \beta=\vja$ and the eigenvalue condition becomes
$\det ( M + N H_{j,k}(1) ) = 0$. A simple example of a set of functions
satisfying (\ref{simini}) are those for the equation with $R_{j}=0$. They are
given by $\ujxe = \cos kx$ and $\ujxz = k^{-1} \sin kx$.

The solution ${\bf u}_{j,k}(x)$, by contrast, depends on the nature of the
boundary conditions. For Robin boundary conditions, for example,
$A\alpha + B\beta = 0$, and we may take $\alpha = -B$ and $\beta = A$,
$\ujx$ being only defined up to a constant as usual. Thus
\beq
{\bf u}_{jk}(x) = -B {\bf u}_{j,k}^{(1)}(x) + A {\bf u}_{j,k}^{(2)}(x) .
\label{Robin_ex}
\eeq
It is now straightforward to calculate $\det (M + NH_{j,k}(1) )$ using the
matrices appropriate for Robin conditions:
\beq
M=\left(
\begin{array}{cc}
A & B \\
0 & 0
\end{array}
\right) , \quad
N=\left(
\begin{array}{cc}
0 & 0 \\
C & D
\end{array}
\right) , \label{MN_Rob}
\eeq
to find that
\beq
\det (M + NH_{j,k}(1) ) = C \ujb + D \vjb , \nn
\eeq
so that imposing the boundary conditions at the ``final'' point gives the
eigenvalue condition (\ref{nontri1}).

For non-separated boundary conditions, initial and final conditions are mixed
together and the appropriate linear combination (\ref{lin_comb}) is determined
by imposing either one of the boundary conditions. For example in the case of
periodic boundary conditions, $\uja = \ujb$ implies that
$\alpha (1-u_{j,k}^{(1)}(1) ) = \beta u_{j,k}^{(2)}(1)$, so that we may take
\beq
\ujx = u_{j,k}^{(2)}(1) \ujxe + \left( 1 - u_{j,k}^{(1)}(1) \right) \ujxz .
\label{period_ex}
\eeq
The other boundary condition then gives the eigenvalue condition. Explicitly
calculating $\det ( M + N H_{j,k}(1) )$ using (\ref{period_ex}) and
$M=-N=I_{2}$, where $I_2$ is the $2 \times 2$ unit matrix, one finds
\beq
\det ( M + N H_{j,k}(1) ) = v_{j,k}(0) - v_{j,k}(1) .
\nn
\eeq

It is now possible to see how the approach adopted to investigate the case
of Dirichlet boundary conditions fits into this more general structure. The
solution ${\bf u}_{j,k}(x)$ which satisfies (\ref{first_order}) is simply
equal to $\ujxz$ introduced in this section, that is, $\alpha=0, \beta=1$.
Using (\ref{MN_Dir}), $\det (M + N H_{j,k}(1) )= u_{j,k}^{(2)}(1)$, so that
the eigenvalues are fixed by imposing $\ujb = 0$, just as in
(\ref{eigencond_Dir}). For Neumann boundary conditions $\alpha=1, \beta=0$; for
other cases both of the fundamental solutions $\ujxe$ and $\ujxz$ will be
required to construct a solution $\ujx$ which has the correct properties.

The discussion presented in this section so far indicates how our
earlier proof can be extended to more general types of boundary
conditions. Assuming suitable analyticity properties in $k$, the 
eigenvalue condition (\ref{eigencond_Dir}) is now
replaced by the more general condition (\ref{nontri1}), and so the
analogue of (\ref{2}) becomes \beq \zeta _{L_1} (s) -\zeta_{L_2}
(s)
       = \frac 1 {2\pi i} \int_\gamma dk k^{-2s}
 \tk \ln \frac{\det (M + N H_{1,k}(1) )}{\det (M + N H_{2,k}(1) )} .
\label{beginfrom} \eeq We next perform the same steps as in
section 2. The leading $|k|\to \infty$ behaviour of $u_{j,k} ^{(2)}
(x)$ is governed by eq. (\ref{CHas}), whereas for $u_{j,k}^{(1)}
(x)$ we have \cite{cou53} 
\beq 
u_{j,k}^{(1)} (x) \sim \cos (kx) (1+ {\cal O} (k^{-1}) ) .
\nn 
\eeq 
This shows 
\beq \frac d {dk} \ln \frac{\det (M+N H_{1,k} (1))}{\det (M+N H_{2,k} (1))} 
= {\cal O}(k^{-2}) 
\nn 
\eeq 
and all goes through as before. Thus, with the definition 
\beq Y_j (x) \equiv \left( \begin{array}{cc}
y_{j}^{(1)} (x) & y_{j}^{(2)} (x) \\
y_{j}^{(1)\prime}(x) & y_{j}^{(2)\prime}(x)
\end{array}
\right)
=
\lim _{k\to 0} H_{j,k} (x) \nn
\eeq
the formalism immediately provides the answer for the quotient
of determinants
\beq
\frac{\det L_1}{\det L_2} = \frac{\det (M +N Y_1 (1) )}
        {\det (M +N Y_2 (1) ) } .
\label{pregen}
\eeq 
From a practical point of view, in order to calculate
(\ref{pregen}) only the matrices $M$ and $N$, which specify the
boundary conditions, and the two independent solutions $\{
y_{j}^{(1)}(x),$ $ y_{j}^{(2)}(x) \}$ of the homogeneous equation
$L_{j} y_{j} = 0$ are needed. These fundamental solutions also
satisfy the initial conditions (\ref{simini}) inherited from the
solutions $\{ u_{j,k}^{(1)}(x), u_{j,k}^{(2)}(x) \}$: 
\beq 
Y_{j} (0) = \left(
\begin{array}{cc}
y_{j}^{(1)} (0) & y_{j}^{(2)} (0) \\
y_{j}^{(1)\prime} (0) & y_{j}^{(2)\prime} (0)
\end{array}
\right)
=
\left(
\begin{array}{cc}
1 & 0 \\
0 & 1
\end{array}
\right) . \label{simini_y}
\eeq
For instance, in the example introduced in section 3 where $R_{1}(x)=x$,
the fundamental solutions are
\beq
y_{1}^{(1)}(x) &=& \frac{3^{-1/3}\,\pi}{\Gamma(1/3)}
\left[ {\rm Bi}(x) + \sqrt{3} {\rm Ai}(x) \right] \nn \\
y_{1}^{(2)}(x) &=& \frac{\Gamma(1/3)}{2\,3^{1/6}}
\left[ {\rm Bi}(x) - \sqrt{3} {\rm Ai}(x) \right] .
\label{fund_airy}
\eeq
For Dirichlet boundary conditions $y_{1}(x)$ is simply $y_{1}^{(2)}(x)$
(compare with (\ref{det_airy0})), and for Neumann boundary conditions it is
$y_{1}^{(1)}(x)$.

Let us end this section by giving an alternative, slightly simpler, form for
(\ref{pregen}) which will be useful in the next section. We have seen in
the case of Robin and periodic boundary conditions, that if we implement
one of the boundary conditions, then $\det ( M + N H_{j,k}(1))$ is proportional
(equal, with the correct choice of normalisation) to the other condition.
This will be true in general. To explicitly show this, let us write the
condition (\ref{gbc}) out in full:
\beq
\left(
\begin{array}{cc}
m_{11} \uja +m_{12} \vja +n_{11} \ujb +n_{12} \vjb \\
m_{21} \uja +m_{22} \vja +n_{21} \ujb +n_{22} \vjb
\end{array}
\right)
=
\left(
\begin{array}{cc}
0 \\
0
\end{array}
\right) .
\label{explicit}
\eeq
We ask that $\ujx$ satisfy only one boundary condition. This could be either
one, but one choice gives
\beq
m_{11} \uja +m_{12} \vja +n_{11} \ujb +n_{12} \vjb = 0.
\label{first_half}
\eeq
We now want to construct a $\ujx$ which explicitly satisfies the condition
(\ref{first_half}). To do so we expand $u_{j,k}(x)$ as in (\ref{lin_comb})
and implement (\ref{first_half}). This gives (with an appropriate choice of
normalisation) \beq \ujx & = & -\left[ m_{12} + n_{11}
u_{j,k}^{(2)}(1) +
n_{12} v_{j,k}^{(2)}(1) \right] \ujxe \nn \\
& + & \left[ m_{11} + n_{11} u_{j,k}^{(1)}(1) + n_{12}
v_{j,k}^{(1)}(1) \right] \ujxz . \label{gen_u} \eeq Using the
expression (\ref{gen_u}) to calculate an explicit expression for
the second row of the left-hand side of (\ref{explicit}) gives
\beq
\det M + \det N &+& u_{j,k} ^{(1)} (1) (n_{11} m_{22} -
n_{21} m_{12} )
+ u_{j,k} ^{(2)} (1) ( m_{11} n_{21} - m_{21} n_{11} ) \nn\\
&+& v_{j,k} ^ {(1)}(1) ( n_{12} m_{22} - n_{22} m_{12}) + v_{j,k}
^{(2)} (1) ( m_{11} n_{22} - m_{21} n_{12} ) ,
\nn
\eeq
which is simply $\det (M + N H_{j,k}(1))$ expanded out and simplified using
$\det H_{j,k}(1) = 1$. Therefore we have shown that
\beq
\det (M+N H_{j,k} (1)) = m_{21} \uja +m_{22} \vja +n_{21} \ujb +n_{22}
\vjb .
\label{detlow}
\eeq
The fact that the constant of proportionality between the two sides of
this equation is 1 is a consequence of the judicious choice of normalisation
of (\ref{gen_u}). If we had chosen to implement the second row of
(\ref{explicit}) instead of (\ref{first_half}), we would simply
have obtained a different $\ujx$, which, when substituted into the
first row of (\ref{explicit}), would have again given $\det (M + N
H_{j,k}(1) )$.

As a result of (\ref{detlow}) we find for the quotient of functional
determinants
\beq
\frac{\det L_1}{\det L_2} = \frac{m_{21} y_1 (0) + m_{22} y_1 ' (0)
  + n_{21} y_1 (1) + n_{22} y_1 ' (1) }
{ m_{21} y_2 (0) + m_{22} y_2 ' (0)
  + n_{21} y_2 (1) + n_{22} y_2 ' (1) },
\label{fungenres}
\eeq
which is the alternative form to (\ref{pregen}) mentioned above. It is now
possible to immediately write down the required formulae for any boundary
condition. For instance, in the case of Robin boundary conditions defined by
(\ref{MN_Rob}),
\beq
\frac{\det L_1}{\det L_2} = \frac{C y_1 (1) + D y_1 ' (1) }
{C y_2 (1) + D y_2 ' (1) },
\label{det_Rob}
\eeq
and for periodic boundary conditions defined by $M = -N = I_2$,
\beq
\frac{\det L_1}{\det L_2} = \frac{y_1 ' (0) - y_1 ' (1) }
{y_2 ' (0) - y_2 ' (1) } .
\label{det_period}
\eeq

\bigskip

\section{Omitting the zero mode: general case}
\label{general_zeromode}

In this section we will derive the result for the ratio of
functional determinants when a zero mode has been extracted, in the
case when general boundary conditions of the type (\ref{gbc}) have
been imposed on operators of the type (\ref{start1}). As before,
we will assume that the operator $L_j$ with the boundary
conditions imposed defines a self-adjoint operator. This requires
that the elements of the matrices $M$ and $N$ satisfy certain conditions,
as described in Appendix A. The main result which we will require
will be the nature of the generalisation of (\ref{4}), where
$u_{1,k}(1)$ will presumably be replaced by $\det ( M + N
H_{j,k}(1) )$. As in section \ref{zeromode}, the starting point of
the analysis is \beq \left[ v_{1,0}(x)^* u_{1,k} (x) - v_{1,k}(x)
u_{1,0} (x)^* \right] ^1_0 = k^2 \langle u_{1,0} | u_{1,k} \rangle
.\label{gen1} \eeq We implement both boundary conditions on the
$k=0$ solution, but only one of the boundary conditions on the
solution for general $k$, since we wish to keep $k$ arbitrary, and
not restrict it so that $k^2$ is an eigenvalue.

To gain some insight, we first look at particular boundary
conditions before we outline the procedure for general boundary
conditions. For Robin boundary conditions we have
$Au_{1,0}(0)+Bv_{1,0}(0)=0, Cu_{1,0}(1)+Dv_{1,0}(1)=0$ and
$Au_{1,k}(0)+Bv_{1,k}(0)=0$. The left-hand side of (\ref{gen1})
then gives zero for the contribution at $x=0$ (both functions
satisfy the boundary conditions there) and at $x=1$ gives \beq
-\frac{C}{D} u_{1,0}(1)^* u_{1,k}(1) - v_{1,k}(1)u_{1,0}(1)^* = -
\left( \frac{u_{1,0}(1)^*}{D} \right) \left[ Cu_{1,k}(1) +
Dv_{1,k}(1) \right]. \nn \eeq Therefore \beq Cu_{1,k}(1) +
Dv_{1,k}(1) = - \left( \frac{D}{u_{1,0}(1)^*} \right)\, k^{2}
\langle u_{1,0}|u_{1,k}\rangle . \label{r_zero} \eeq For periodic
boundary conditions we have $u_{1,0}(0)=u_{1,0}(1),
v_{1,0}(0)=v_{1,0}(1)$ and $u_{1,k}(0)=u_{1,k}(1)$. The left-hand
side of (\ref{gen1}) then gives \beq -v_{1,k}(1)u_{1,0}(0)^* +
v_{1,k}(0)u_{1,0}(0)^* \nn \eeq which leads to \beq v_{1,k}(0) -
v_{1,k}(1) = \frac{1}{u_{1,0}(0)^*}\,k^{2}\langle u_{1,0}| u_{1,k}
\rangle . \label{p_zero} \eeq For all special examples considered,
the result takes the form (see eqs. (\ref{4}), (\ref{r_zero}),
(\ref{p_zero})) \beq \det (M+N H_{1,k} (1) ) = {\cal B} k^2
\langle u_{1,0} | u_{1,k} \rangle ,\label{answergen} \eeq where
${\cal B}$ depends on the zero mode data at the boundary and on
the boundary condition imposed.

In fact, the procedure described above can be extended to the case of
general boundary conditions and the analogous results will always have
the form (\ref{answergen}). To see this recall from the last section that
implementing one boundary condition only leads, in the case of general
boundary conditions, to (see (\ref{first_half}) and (\ref{detlow}))
\beq
m_{11} u_{1,k} (0) + m_{12} v_{1,k} (0) + n_{11} u_{1,k} (1)
+ n_{12} v_{1,k} (1) &=& 0 \label{gen2}\\
m_{21} u_{1,k} (0) + m_{22} v_{1,k} (0) + n_{21} u_{1,k} (1)
+ n_{22} v_{1,k} (1) &=& \drel .\nn
\eeq
When $k=0$, $u_{1,k}(x)$ is the zero mode and so (\ref{gen2}) holds, but with
$\det (M + N H_{1,0}(1))$ replaced by 0.

We wish to obtain expressions for $u_{1,k}(x), u_{1,0}(x)$, and
their derivatives, at the boundaries in order to substitute them
into the left-hand side of (\ref{gen1}), and ultimately obtain an
expression for $\drel$. To do this we have to proceed by cases,
which depend on the particular boundary conditions prescribed.
Suppose that, for instance, the boundary conditions are such that
$m_{12} n_{22} - n_{12} m_{22} \neq 0$. Then in this case we
rewrite (\ref{gen2}) as \beq \left(
\begin{array}{cc}
m_{12} & n_{12} \\
m_{22} & n_{22}
\end{array}
\right)
{v_{1,k} (0) \choose v_{1,k} (1) }
=
{0 \choose \drel } -
\left(
\begin{array}{cc}
m_{11} & n_{11} \\
m_{21} & n_{21}
\end{array}
\right) {u_{1,k} (0) \choose u_{1,k} (1) } . \label{first_case}
\eeq The condition on the elements of $M$ and $N$ that we have
imposed implies that the matrix on the extreme left of
(\ref{first_case}) is invertible and so we may solve for
$v_{1,k}(0)$ and $v_{1,k}(1)$. These intermediate results are of
no intrinsic interest and are given in the Appendix A. Obviously
the equations for $v_{1,0}(0)$ and $v_{1,0}(1)$ are identical to
those for general $k$, except that $\drel$ is replaced by $\det (M
+ N H_{1,0}(1)) = 0$. We may now eliminate $v_{1,k}(0),
v_{1,k}(1), v_{1,0}(0)$ and $v_{1,0}(1)$ from (\ref{gen1}) using
these equations. The basic steps are outlined in Appendix A. We
obtain 
\beq 
\drel = \frac{n_{12} m_{22} - m_{12} n_{22}}{
 m_{12} u_{1,0} (1)^* + n_{12} u_{1,0} (0) ^*}
\langle u_{1,0} | u_{1,k} \rangle, \quad
n_{12} m_{22} - m_{12} n_{22} \neq 0 .
\label{detgen1}
\eeq
Expressing $\{ v_{1,k} (0), v_{1,k} (1) \}$ in terms of
$\{ u_{1,k} (0), u_{1,k} (1) \} $ is just one possibility; there are five
more ways to express two boundary data by the complementary ones. Proceeding
as described gives the list of results presented in Appendix A. It is seen
that the precise form that ${\cal B}$ takes depends on which combination of
elements of $M$ and $N$ are non-zero. The formulae in Appendix A, together
with (\ref{detgen1}), cover all possible cases. If more than one of the
results is appropriate, then they are, of course, equivalent.

Having established that (\ref{answergen}) holds in general, we may simply
replace (\ref{4}) by
\beq
\drel = - k^{2}f_{1,k} ; \ \ f_{1,k} \equiv - {\cal B}
\langle u_{1,0} | u_{1,k} \rangle ,
\label{gen_zm_res}
\eeq
and repeat the steps in section \ref{zeromode}. One finds
\beq
\zeta ' _{L_1} (0) - \zeta ' _{L_2 } (0) =
-\ln  \frac{f_{1,0}}{\det ( M + N Y_{2}(1) )} , \nn
\eeq
which using (\ref{det_defn}) gives
\beq
\frac{\det ' L_{1}}{\det L_{2}} = -
\frac{ {\cal B}\,\langle y_{1} | y_{1} \rangle}{\det ( M + N Y_{2}(1) )} ,
\label{primed2}
\eeq
where ${\cal B}$ takes on the following values:
\beq
\frac{n_{12} m_{22} - m_{12} n_{22}}{
 m_{12} y_{1} (1)^* + n_{12} y_{1} (0)^* } , \quad
{\rm if} \ \ n_{12} m_{22} - m_{12} n_{22} &\neq& 0 ,
\nn\\
\frac{m_{11} n_{21} - m_{21} n_{11}}{
 m_{11} y_{1} ' (1)^* + n_{11} y_{1} ' (0)^* } , \quad
{\rm if} \ \ m_{11} n_{21} - m_{21} n_{11} &\neq& 0 ,
\nn\\
\frac{m_{11} m_{22} - m_{12} m_{21}}{
 m_{12} y_{1} ' (0)^* + m_{11} y_{1} (0) ^*} , \quad
{\rm if} \ \ m_{11} m_{22} - m_{12} m_{21} &\neq& 0 ,
\nn\\
\frac{m_{12} n_{21} - m_{22} n_{11}}{
 m_{12} y_{1} '  (1)^* - n_{11} y_{1} (0) ^*} , \quad
{\rm if} \ \ m_{12} n_{21} - m_{22} n_{11} &\neq& 0 ,
\nn\\
\frac{n_{12} n_{21} - n_{22} n_{11}}{
 n_{12} y_{1} ' (1)^* + n_{11} y_{1} (1)^* } , \quad
{\rm if} \ \ n_{12} n_{21} - n_{22} n_{11} &\neq& 0 ,
\nn\\
\frac{m_{11} n_{22} - m_{21} n_{12}}{
 n_{12} y_{1} ' (0)^* - m_{11} y_{1} (1)^* } , \quad
{\rm if} \ \ m_{11} n_{22} - m_{21} n_{12} &\neq& 0 .
\label{B}
\eeq

As an example, for periodic boundary conditions when $M=I_2$, the
third of the above results is applicable, and therefore ${\cal B}
= 1/y_{1} (0)^{*}$. Equivalently the fifth result may be used, since
$N=-I_2$, to give ${\cal B} = 1/y_{1} (1)^{*}$. Furthermore $\det
(M+NY_{2}(1))=y_{2}' (0)-y_{2}' (1)$ (see (\ref{det_period})), so
that
\beq
\frac{\det ' L_{1}}{\det L_{2}} = - \frac{ \langle y_{1}
| y_{1} \rangle }{y_{1}(0)^*\,\left[ y_{2}' (0) -y_{2}' (1)
\right] } .
\label{detprimed_period}
\eeq

Let us end this section by showing that (\ref{detprimed_period})
is in agreement with the result obtained by a different
method~\cite{mck95}. In order to show that the two results are
identical, we need to establish a correspondence between the
solutions utilised in the two approaches. In this paper we have
taken $\ujxe$ and $\ujxz$ --- which in the $k \to 0$ limit become
$y^{(1)} (x)$ and $y^{(2)} (x)$ --- as the two independent
solutions (to facilitate comparisons with the formulae given in
Ref. \cite{mck95}, we will drop the $j$ subscript for the rest of
this section). In Ref. \cite{mck95} the two independent solutions
were denoted by $y_{1} (x)$ and $y_{2} (x)$, but we will denote
them here by $y^{({\rm I})} (x)$ and $y^{({\rm II})} (x)$. The
solution $y^{({\rm I})} (x)$ is simply what we have called $y (x)$
in this paper, that is, the solution which is the zero mode if one
exists, and which satisfies only one of the boundary conditions,
if a zero mode does not exist. Thus from (\ref{lin_comb}), when $k
\to 0$, \beq \left(
\begin{array}{cc}
y^{({\rm I})} (x) & y^{({\rm II})} (x) \\
y^{({\rm I})\prime} (x) & y^{({\rm II})\prime} (x)
\end{array}
\right)
=
\left(
\begin{array}{cc}
y^{(1)} (x) & y^{(2)} (x) \\
y^{(1)\prime} (x) & y^{(2)\prime} (x)
\end{array}
\right)
\left(
\begin{array}{cc}
\alpha & \gamma \\
\beta & \delta
\end{array}
\right) ,
\label{mapping}
\eeq
where $y^{({\rm II})} (x) = \gamma y^{(1)} (x) + \delta y^{(2)} (x)$ is any
solution which is independent of $y^{({\rm I})} (x)$. Since $\det Y (x) = 1$
in our formalism, the Wronskian of the two solutions $y^{({\rm I})} (x)$ and
$y^{({\rm II})} (x)$ is
\beq
y^{({\rm I})} (x) y^{({\rm II})\prime} (x) -
y^{({\rm II})} (x) y^{({\rm I})\prime} (x) = \alpha \delta - \beta \gamma .
\label{MT_W}
\eeq

All of this holds whatever the nature of the boundary conditions
--- they will correspond to particular choices for $\alpha$ and
$\beta$. Specifically, for periodic boundary conditions,
(\ref{period_ex}) shows that $\alpha = y^{(2)} (1)$ and $\beta = 1
- y^{(1)} (1)$, so that the Wronskian, W, is given by $W = \delta
y^{(2)} (1)-\gamma \left( 1 - y^{(1)} (1) \right)$. It is
straightforward to show that $y^{({\rm II})} (1) - y^{({\rm II})}
(0)$ is also equal to this quantity. So, for periodic boundary
conditions, and $y^{({\rm I})} (x)$ given by the $k \to 0$ version
of (\ref{period_ex}), $W = y^{({\rm II})} (1) - y^{({\rm II})}
(0)$. This verifies that eq. (5.2) of Ref. \cite{mck95} reduces to
(\ref{detprimed_period}) above, up to a minus sign which
originates from an overall sign difference in the definition of
the operators $L_j$.

\bigskip

\section{Systems of differential operators}
\label{systems}

As we have stressed several times already, our main aim in this paper is to
present results and proofs in such a way as to be as free of technical details
as possible. So, while the results for a second order differential operator
of the type (\ref{start1}) presented in the earlier sections of the paper
generalise to systems of differential operators, we will only give a brief
outline of the formalism here, leaving the details to a forthcoming
publication~\cite{kir03}. In particular, we will, for definiteness, discuss
most of the results in the context of a specific example, encountered in the
study of transition rates between metastable states in superconducting
rings~\cite{tar94,tar95}. The differential operator in this problem is defined
on the interval $[-l/2, l/2]$ and has the form
\beq
L_{1} =
\left(
\begin{array}{cc}
- \frac{d^{2}}{dx^{2}} + (1-2\mu^{2}) & (1-\mu^{2})e^{2i\mu x} \\ \\
(1-\mu^{2})e^{-2i\mu x} & - \frac{d^{2}}{dx^{2}} + (1-2\mu^{2})
\end{array}
\right) \equiv - I_{2}\,\frac{d^2}{dx^2} + Q_{1} (x).
\label{start2}
\eeq
In the actual problem this example is taken from, $\mu$ is a wavevector,
which is small (so that $\mu^{2} < 1$), and $l$ is the circumference of the
ring divided by the temperature-dependent correlation length. The solutions
of $\left( L_{j} - k^{2} \right) {\bf u}_{j,k} (x)=0$ will be denoted
explicitly by
\beq
{\bf u}_{j,k} (x) = {u_{j,k,1} (x) \choose u_{j,k,2} (x) } .
\label{bold_u_def}
\eeq
The sole purpose of the operator $L_2$ is to render the ratio
$\det L_{1}/\det L_{2}$ finite, and it is chosen on grounds of simplicity
and convenience.

In the completely general case ${\bf u}_{j,k} (x)$ will have $r$
components $u_{j,k,a} (x) ; a=1,\ldots,r$, and $L_j$ will take the
form of an $r \times r$ matrix containing second order derivatives
along the diagonal. It is again convenient to go over to the
equivalent problem containing only first order derivatives, which
will have the form (\ref{first_order}), but with the column
vectors having entries $u_{j,k,a} (x)$ and $v_{j,k,a} (x);
a=1,\ldots,r$, and with the square matrix having the structure
\beq \left( \begin{array}{c|c}
  0 & I_{r} \\[.1cm] \hline\\[-.2cm]
 Q_{j} (x) - k^{2}I_{r} & 0 \end{array} \right) ,
\label{square}
\eeq
where $I_r$ is the $r \times r$ unit matrix and $Q_j (x)$ contains the
non-derivative terms in $L_j$, as in our specific example (\ref{start2}). The
boundary conditions can again be expressed as (\ref{gbc}), but now with
$M$ and $N$ being $2r \times 2r$ matrices. The advantage of defining the
structure in this way is that much of the formalism described in
section \ref{general} goes through unchanged. For instance, the fundamental
solution consists of $2r$ independent solutions of $L_1$ and their derivatives:
\beq
H_{j,k} (x) =
\left(
\begin{array}{cccc}
u_{j,k,1}^{(1)} (x)  & \ldots & \ldots & u_{j,k,1}^{(2r)} (x) \\
  \ldots             & \ldots & \ldots & \ldots \\
u_{j,k,r}^{(1)} (x) & \ldots & \ldots & u_{j,k,r}^{(2r)} (x) \\ \\
v_{j,k,1}^{(1)} (x)  & \ldots & \ldots & v_{j,k,1}^{(2r)} (x) \\
  \ldots             & \ldots & \ldots & \ldots \\
v_{j,k,r}^{(1)} (x) & \ldots & \ldots & v_{j,k,r}^{(2r)} (x)
\end{array}
\right) ,
\label{rbyrH}
\eeq
and the eigenvalue condition is again
(\ref{nontri1}). We again choose the solutions so that
$H_{j,k}(-l/2) = I_{4}$, just as we did in the one-component case
(\ref{simini}). When no zero modes are present, the representation
(\ref{beginfrom}) then leads to the final result (\ref{pregen}) as
before. To calculate the right-hand side of (\ref{pregen}) we only
need solutions of the homogeneous equations $L_{j}{\bf y}_{j}=0$
and the matrices $M$ and $N$.

In the particular example we are considering here, we will impose
``twisted boundary conditions'' \cite{tar94,tar95} $M = - {\rm
diag} \left\{ e^{i\mu l}, e^{-i\mu l}, e^{i\mu l}, e^{-i\mu l}
\right\}$ and $N=I_4$.
It is easy to check that
\beq
{\bf y}_{1} (x) = { e^{i\mu x} \choose - e^{-i\mu x} } ,
\label{zm}
\eeq
is an eigenfunction of the problem with zero eigenvalue and so
$\det (M + N H_{1,0} (l/2)) = 0$.

Since a zero mode is present we have to proceed in the general way
indicated in section \ref{general_zeromode}, but suitably
generalised to this particular two-component case. The analog of
({\ref{gen1}) is
\beq
\sum^{2}_{a=1}\,\left[ v_{1,0,a}^{*} (x)
u_{1,k,a} (x) - v_{1,k,a} (x) u_{1,0,a}^{*} (x)
\right]^{l/2}_{-l/2} = k^{2} \langle {\bf u}_{1,0} | {\bf u}_{1,k}
\rangle .
\label{gen1_2cpt}
\eeq
Now we assume that ${\bf u}_{1,0}(x)$ satisfies all of the boundary
conditions, so it is a zero mode, but ${\bf u}_{1,k} (x)$ satisfies
only 3 of the 4 --- for instance, those used in Appendix B, see
(\ref{bcsup}). This means that $k$ is not restricted to values for
which $k^2$ is an eigenvalue. Substituting in these conditions,
(\ref{gen1_2cpt}) becomes
\beq
- \left[ v_{1,k,2} (l/2) - e^{-i\mu l} v_{1,k,2}
(-l/2) \right] u_{1,0,2}^{*} (l/2) = k^{2} \langle {\bf u}_{1,0} |
{\bf u}_{1,k} \rangle .
\label{gen2_2cpt}
\eeq
Only the component of ${\bf u}_{1,k} (x)$ that does not satisfy the boundary
condition survives on the left-hand side of the equation. If
${\bf u}_{1,k} (x)$ is appropriately normalised, the term in square brackets
on the left-hand side of (\ref{gen2_2cpt}) is $\det ( M + NH_{1,k} (l/2) )$,
and we find
\beq
\det ( M + NH_{1,k} (l/2) ) = - \frac{k^{2}}{u_{1,0,2}^{*}
(l/2)}\, \langle {\bf u}_{1,0} | {\bf u}_{1,k} \rangle  \equiv - k^{2}
f_{1,k} ,
\label{gen3_2cpt}
\eeq
where $f_{1,k} = - {\cal B}
\langle {\bf u}_{1,0} | {\bf u}_{1,k} \rangle$ and ${\cal B} = -
\left\{ y_{1,2}^{*} (l/2) \right\}^{-1}$. The only remaining task is
the calculation of the appropriately normalised function ${\bf y}_1 (x)$.
The method used to obtain it is outlined in Appendix B. Using this procedure
we find
\beq
{\bf y}_{1} (x) = - \frac{4 l (1-\mu^{2})
\sinh^{2}\left( \frac{l\nu}{2} \right) e^{i l \mu /2} }{\nu^2} { e^{i\mu x}
\choose - e^{-i\mu x} } ,
\label{zm_norm}
\eeq
which should be contrasted with the unnormalised result (\ref{zm}). Therefore
\beq
f_{1,0} = \frac{\langle {\bf y}_{1} | {\bf y}_{1} \rangle}{y_{1,2}^*(l/2)}
= \frac{ 8 l^{2} (1-\mu^2) \sinh ^2 \left(\frac{ l \nu } 2 \right) } { \nu ^2},
\label{f1k_2cpt}
\eeq
which agrees with the result calculated in Ref.~\cite{tar95} using a different
method. This result gives $\det ' L_1$, provided that we divide by $\det L_2$.
The most convenient choice for $L_2$ is to take it to be of the form
$L_2 = - I_2 (d^2/dx^2)$. The calculation of $\det L_2$ then decouples into 
the calculation of the determinant of two identical operators of the kind  
studied in the earlier sections of this paper.

\bigskip

\section{Conclusion}
\label{conclusion}

The two main objectives of this paper have been firstly to present a simple,
yet rigorous, way of determining the now well-established formulae for ratios
of functional determinants, and secondly to extend the results to the
situation where a zero mode exists, but is omitted, in one of the
determinants. We have stressed clarity, preferring to study the simplest
types of operators, and leaving the most general case to a later
publication~\cite{kir03}.

The main idea behind the method of proof which we have adopted
involved defining two solutions of the eigenvalue equation
$L_{j}\ujx = k^{2} \ujx$, denoted by $\ujxe$ and $\ujxz$. Here
$L_1$ is the second order differential operator of interest, and
$L_2$ is a similar, typically simpler, operator used for
normalisation. Although $\ujxe$ and $\ujxz$ satisfy the eigenvalue
equation, they are not, in general, eigenfunctions since they do
not satisfy the boundary conditions (simple boundary conditions
such as Dirichlet and Neumann are exceptions). Instead we impose
``initial'' conditions on these two solutions --- conditions on
the functions and their first derivatives at $x=0$ --- so that
they are uniquely defined. The particular choice we made in this
paper was (\ref{simini}), but obviously other choices are
permissible.

Once two unique and independent solutions of the eigenvalue equation have been
defined, any other solution may be expressed as a linear combination of these
by specifying the two constants $\alpha$ and $\beta$, as in (\ref{lin_comb}).
In particular, if we wish to determine eigenfunctions we would first apply one
of the boundary conditions to find the ratio of $\alpha$ to $\beta$. Since
the overall normalisation of eigenfunctions is irrelevant, this application
of only one of the boundary conditions determines the functional form of the
eigenfunction $\ujx$. The imposition of the second boundary condition has
the effect of restricting values of $k^2$ to those which are
eigenvalues --- which are real, since the operators considered here are all
taken to be self-adjoint.

The proof of the results begins from the observation that the quantity
$\det ( M + N H_{j,k} (1) )$ is zero only when $k^2$ is an eigenvalue. It
follows that its logarithmic derivative has poles with unit residue at values
of $k$ such that $k^2$ is an eigenvalue. Therefore a zeta-function can be
defined through a contour integration, where the contour is chosen to surround
these values of $k$. Since these poles are restricted to the non-negative
real axis, the contour may be deformed to lie on the imaginary $k$ axis, and
then evaluated to give the required result (\ref{pregen}). The only other
information which is required during this process relates to the nature of
the solutions of the differential equation for large $|k|$~\cite{cou53}.

An alternative result, which is slightly more explicit than (\ref{pregen}),
can be obtained by using the solution
$y_{j} (x) \equiv \lim_{k \to 0} \ujx$ rather
than $y^{(1)}_{j} (x) \equiv \lim_{k \to 0} \ujxe$ and
$y^{(2)}_{j} (x) \equiv \lim_{k \to 0} \ujxz$, to express the result. Here
one has to be careful with questions of normalisation. The solution $\ujx$
satisfies only one of the boundary conditions, but is unnormalised. However
if its normalisation is chosen appropriately, then
$\det ( M + N H_{j,k} (1) )$ can be made to exactly equal the second boundary
condition (that is, with a constant of proportionality equal to 1). The
effect of this, after the $k \rightarrow 0$ limit has been taken, is to allow
$\det ( M + N H_{j,0} (1) )$ in (\ref{pregen}) to be replaced by the second
boundary condition, where the function appearing in this boundary condition
is simply $y_{j} (x)$, appropriately normalised. In the case of the
generalised boundary conditions considered in section \ref{general}, the
appropriately normalised $\ujx$ satisfying the first part of the boundary
conditions is given by (\ref{gen_u}), the expression of
$\det ( M + N H_{j,k} (1) )$ in terms of the second boundary condition is
given by (\ref{detlow}), and the final result, in which the ratios of
functional determinants are expressed in terms of $y_{j} (x)$, is
(\ref{fungenres}).

One of the advantages of the above method of proof is that it can
be relatively simply adapted to the case when $k=0$ is an
eigenvalue of $L_1$ which is excluded from the determinant. This
is achieved by noting that in these cases $\det ( M + N H_{1,k}
(1)) = - k^{2} f_{1,k}$, where $f_{1,0} \neq 0$. Since $f_{1,k}$
will, like $\det ( M + N H_{1,k} (1))$, vanish at all the non-zero
eigenvalues, the proof we have described can be repeated but with
$\det ( M + N H_{1,k} (1) )$ replaced by $f_{1,k}$ --- or rather,
by $(1-k^{2}) f_{1,k}$, so that the large $|k|$ behaviour remains
unchanged. The result will be a formula for the ratio of
functional determinants, but with the zero mode omitted in the
determinant in the numerator. It is given by (\ref{primed2}), and
although it depends on a constant ${\cal B}$ whose form depends on
the nature of the boundary conditions, it is nonetheless simple,
in the sense that it only depends on (i) the zero mode $y_{1}
(x)$, and (ii) the matrices $M$ and $N$ which define the boundary
conditions. Furthermore, we expect that in many applications the
norm $\langle y_{1} | y_{1} \rangle$ will cancel with a similar
factor resulting from the transformation to a set of coordinates
which allows easy extraction of the zero mode. So, once again,
only the values of solutions of the homogeneous equation at the
boundary are required.

In the examples given in this paper, we have been able to obtain the solutions
of the homogeneous equations analytically, and so determine their values,
and/or those of their derivatives, at the boundaries. In most cases of
interest this will not be possible, but this does not create any real
difficulty: it is straightforward enough to determine the solutions
$y^{(1)}_{j} (x)$ and $y^{(2)}_{j} (x)$ numerically and so calculate the
required result from either (\ref{pregen}) or (\ref{fungenres}). Since
functional determinants of this sort are often perceived as being the most
difficult quantities to evaluate in calculations of fluctuations about
classical field configurations, we hope that the results presented in this
paper will prove useful in this context.

\vspace{1cm}

{\bf Acknowledgements}: We would like to thank Peter Gilkey for
very interesting discussions on the subject. The research of KK
was partially supported by the Max-Planck-Institute for
Mathematics in the Sciences, Leipzig.
\newpage

\renewcommand{\theequation}{\Alph{section}.\arabic{equation}}
\setcounter{section}{0} \setcounter{equation}{0}

\begin{appendix}

\section{Determinants in the presence of zero modes
for general boundary conditions}
\label{appendix1}

In this section we give details of some of the algebraic manipulations used
in section \ref{general_zeromode} and collect together the results for the
different cases.

Let us first consider the case discussed in the main part of the text, that is,
boundary conditions which are such that $m_{12} n_{22} - m_{22} n_{12} \neq 0$.
This condition allows us to rewrite (\ref{first_case}) as
\beq
{v_{1,k} (0) \choose v_{1,k} (1) } &=&
\left(
\begin{array}{cc}
m_{12} & n_{12} \\
m_{22} & n_{22}
\end{array}
\right) ^{-1} {0 \choose \drel } \nn\\ \nn\\
&-&
\left(
\begin{array}{cc}
m_{12} & n_{12} \\
m_{22} & n_{22}
\end{array}
\right) ^{-1}
\left(
\begin{array}{cc}
m_{11} & n_{11} \\
m_{21} & n_{21}
\end{array}
\right)
{u_{1,k} (0) \choose u_{1,k} (1) } .
\nn
\eeq
Multiplying this out yields
\beq
v_{1,k}(0) &=& \frac{1}{\left[m_{12}n_{22} - n_{12}m_{22}\right] }\,
\left\{ -n_{12}\drel \right. \nn\\ \nn \\
&-& \left. \left[ m_{11}n_{22} - n_{12}m_{21} \right] u_{1,k}(0)
- \left[ n_{11}n_{22} - n_{12}n_{21} \right] u_{1,k}(1) \right\}
\nn
\eeq
and
\beq
v_{1,k}(1) &=& \frac{1}{\left[m_{12}n_{22} - n_{12}m_{22}\right] }\,
\left\{ m_{12}\drel \right. \nn\\ \nn \\
&-& \left. \left[ m_{12}m_{21} - m_{11}m_{22} \right] u_{1,k}(0) -
\left[ m_{12}n_{21} - n_{11}m_{22} \right] u_{1,k}(1) \right\} .
\nn
\eeq
Two similar equations hold for $v_{1,0}(0)$ and
$v_{1,0}(1)$, but with $\drel$ replaced by 0. Using these four
equations we find
\beq
\left[ v_{1,0}(x)^* u_{1,k} (x) -
v_{1,k}(x) u_{1,0} (x)^* \right] ^1_0 &=& - \frac{\left[
m_{12}u_{1,0}(1)^*+n_{12}u_{1,0}(0)^*\right] } { \left[
m_{12}n_{22} - n_{12}m_{22} \right] }\,\drel \nn \\
& &\hspace{-3cm} + u_{1,k} (1) u_{1,0} (0)^* \left\{
 - \frac {\det N}{ m_{12} n_{22} - n_{12} m_{22} } +
 \frac{ (\det M ) ^* } { (m_{12} n_{22} - n_{12} m_{22}) ^*}
 \right\} \nn\\
& &\hspace{-3cm} + u_{1,k} (0) u_{1,0} (1)^* \left\{
 - \frac {\det M}{ m_{12} n_{22} - n_{12} m_{22} } +
 \frac{ (\det N ) ^* } { (m_{12} n_{22} - n_{12} m_{22})^* }
 \right\} \nn\\
& &\hspace{-3cm} + u_{1,k} (1) u_{1,0} (1)^* \left\{
 \frac {m_{12} n_{21} - n_{11} m_{22} }{ m_{12} n_{22} - n_{12} m_{22} }
 -\frac{ (m_{12} n_{21} - n_{11} m_{22} ) ^* }
{ (m_{12} n_{22} - n_{12} m_{22})^* } \right\} \nn\\
& &\hspace{-3cm} + u_{1,k} (0) u_{1,0} (0)^* \left\{
  \frac { (m_{11} n_{22} - n_{12} m_{21} ) ^* }
{( m_{12} n_{22} - n_{12} m_{22}) ^* }
  - \frac{ (m_{11} n_{22} - n_{12} m_{21} ) }
{ (m_{12} n_{22} - n_{12} m_{22}) } \right\}
\nn
\eeq
If $k$ is an eigenvalue, the first term in the above expression vanishes.
Therefore, in the case that $k$ is an eigenvalue, in order to have a
self-adjoint problem the sum of the last four terms must vanish. But this
can only happen for general $u_{1,k}$ if each of the four terms in curly
brackets is separately zero. This gives conditions on the entries of the
matrices $M$ and $N$ for the problem to be self-adjoint. We will leave the
discussion of these to a subsequent paper~\cite{kir03}, although let us
note that in the case where $M$ and $N$ are real, these reduce to
$\det M = \det N$.

Therefore, for general $k$, we have using (\ref{gen1}),
\beq
\sca = \frac{\left[m_{12}u_{1,0}(1)^*+n_{12}u_{1,0}(0)^* \right] } { \left[
n_{12}m_{22} - m_{12}n_{22} \right] }\,\drel ,
\nn
\eeq
which yields (\ref{detgen1}).

If, for the particular boundary conditions under consideration,
$m_{12}n_{22} - n_{12}m_{22} = 0$, then (\ref{gen2}) should be expressed not
as (\ref{first_case}), but in an alternative way that allows the above
derivation to be repeated. For example, if
$m_{11}n_{21} - n_{11}m_{21} \neq 0$,
\beq
\left(
\begin{array}{cc}
m_{11} & n_{11} \\
m_{21} & n_{21}
\end{array}
\right)
{u_{1,k} (0) \choose u_{1,k} (1) }
=
{0 \choose \drel } -
\left(
\begin{array}{cc}
m_{12} & n_{12} \\
m_{22} & n_{22}
\end{array}
\right)
{v_{1,k} (0) \choose v_{1,k} (1) } ,
\nn
\eeq
and we can solve for $u_{1,k}(0)$ and $u_{1,k}(1)$. Proceeding in this way
we obtain the following results:
\beq
\drel &=& \frac{m_{11} n_{21} - m_{21} n_{11}}{
 m_{11} v_{1,0} (1)^* + n_{11} v_{1,0} (0)^* } \sca , \quad
m_{11} n_{21} - m_{21} n_{11} \neq 0 ,
\nn\\
\drel &=& \frac{m_{11} m_{22} - m_{12} m_{21}}{
 m_{12} v_{1,0} (0)^* + m_{11} u_{1,0} (0)^* } \sca , \quad
m_{11} m_{22} - m_{12} m_{21} \neq 0 ,
\nn\\
\drel &=& \frac{m_{12} n_{21} - m_{22} n_{11}}{
 m_{12} v_{1,0} (1)^* - n_{11} u_{1,0} (0)^* } \sca , \quad
m_{12} n_{21} - m_{22} n_{11} \neq 0 ,
\nn\\
\drel &=& \frac{n_{12} n_{21} - n_{22} n_{11}}{
 n_{12} v_{1,0} (1)^* + n_{11} u_{1,0} (1)^* } \sca , \quad
n_{12} n_{21} - n_{22} n_{11} \neq 0 ,
\nn\\
\drel &=& \frac{m_{11} n_{22} - m_{21} n_{12}}{
 n_{12} v_{1,0} (0)^* - m_{11} u_{1,0} (1) ^*} \sca , \quad
m_{11} n_{22} - m_{21} n_{12} \neq 0 .\nn
\eeq
Together with (\ref{detgen1}) these equations cover all possible cases.

\bigskip

\section{Explicit calculations for the two-component problem}
\label{appendix2}

In this appendix we will give details of the calculation required to calculate
the functional determinant of the problem defined by (\ref{start2}) on the
interval $[-l/2,l/2]$ with boundary conditions given by the matrices
$M = - {\rm diag} \left\{ e^{i\mu l}, e^{-i\mu l}, e^{i\mu l},
e^{-i\mu l} \right\}$ and $N=I_4$.

The evaluation of the functional determinant $\det ' L_1$ is based
on the evaluation of suitably normalised solutions. In particular
we need to find ${\bf u}_{1,k} (x) = (u_{1,k,1} (x), u_{1,k,2}
(x),$ $v_{1,k,1} (x), v_{1,k,2} (x) )$ such that all but one
boundary condition are satisfied. We choose a normalisation such
that
\beq
(M + N H_{1,k} (l/2) ) \left(
\begin{array}{c} u_{1,k,1} (-l/2)
\\ u_{1,k,2} (-l/2)\\ v_{1,k,1}(-l/2) \\ v_{1,k,2}(-l/2)
 \end{array} \right) = \left(
\begin{array}{c} 0 \\ 0 \\ 0 \\ \det ( M+N H_{1,k} (l/2) )\end{array} \right) .
\label{bcsup}
\eeq
This should be compared with the analogous choices (\ref{first_half}) and
(\ref{detlow}) in section \ref{general}. For the twisted boundary condition
under consideration, this implies
\beq
\det (M+N H_{1,k} (l/2) ) = v_{1,k,2} (l/2) - e^{- i \mu l}
v_{1,k,2} (-l/2),
\label{fourth_cond}
\eeq
which is used in the main text to derive (\ref{gen2_2cpt}). In order to
derive a solution satisfying the condition (\ref{bcsup}) we first note
that when $k$ is not an eigenvalue (so that, in particular, $k \neq 0$), we
can invert this equation to find
\beq
\left(
\begin{array}{c} u_{1,k,1} (-l/2)
\\ u_{1,k,2} (-l/2)\\ v_{1,k,1}(-l/2) \\ v_{1,k,2}(-l/2)
 \end{array} \right) = (M+N H_{1,k} (l/2))^{-1} \left(
\begin{array}{c} 0 \\ 0\\ 0 \\ \det ( M+N H_{1,k} (l/2) ) \end{array} \right).
\label{incon} \eeq Given the fundamental matrix $H_{1,k} (l/2)$,
(\ref{incon}) can be used to find the initial conditions that have
to be imposed on ${\bf u}_{1,k} (x)$ for (\ref{fourth_cond}) to
hold. However, what actually enters into the final answer is
$f_{1,0}$, the $k \to 0$ value of $f_{1,k}$ defined by
(\ref{gen3_2cpt}). Therefore we only need to evaluate
(\ref{incon}) in the limit $k \to 0$. Although $(M+N H_{1,0}
(l/2))^{-1}$ does not exist (because $k=0$ is an eigenvalue) the
right-hand side does exist in the limit, because the determinant
cancels between the inverse and the entry in the column vector
leaving only the cofactors of the matrix which are perfectly well
defined in the $k \to 0$ limit. To calculate these cofactors we
only need the entries of the fundamental matrix $H_{1,k} (l/2)$ at
$k=0$. Thus we only need to construct the normalised fundamental
matrix for $k=0$. Defining, as in section \ref{general}, $Y_{j}
(x) \equiv \lim_{k \to 0} H_{j,k} (x)$, this means we have to
construct \beq Y_{1} (x) \equiv \left(
\begin{array}{cccc}
y_{1,1}^{(1)} (x) & y_{1,1}^{(2)} (x) & y_{1,1}^{(3)} (x) &
y_{1,1}^{(4)} (x)
\\[.1cm]
y_{1,2}^{(1)} (x) & y_{1,2}^{(2)} (x) & y_{1,2}^{(3)} (x) &
y_{1,2}^{(4)} (x)
\\[.1cm]
y_{1,1}^{(1)\prime}(x) & y_{1,1}^{(2)\prime}(x) &
y_{1,1}^{(3)\prime}(x) &
y_{1,1}^{(4)\prime}(x) \\[.1cm]
y_{1,2}^{(1)\prime}(x) & y_{1,2}^{(2)\prime}(x) &
y_{1,2}^{(3)\prime}(x) & y_{1,2}^{(4)\prime}(x)
\end{array}
\right) ,
\nn
\eeq
where the 4 fundamental solutions are chosen to satisfy $Y_1 (-l/2) =I_4$. The
explicit form of these solutions is
\beq
{\bf y}^{(1)}_{1} (x) &=&
\frac{e^{i\mu l/2}}{\nu^{2}}\left(
\begin{array}{c}
\left[ \frac{\nu^2}{2} + i\mu(1-\mu^{2})z + \frac{\nu^2}{2}\cosh \nu z
-\frac{i\mu}{\nu} (3-7\mu^{2})\sinh \nu z \right] e^{i\mu x} \\ \\
(1-\mu^{2}) \left[ - 1 - i\mu z + \cosh \nu z
+\frac{i\mu}{\nu} \sinh \nu z \right] e^{-i\mu x}
\end{array}
\right) \nn\\ \nn\\
{\bf y}^{(2)}_{1} (x) &=&
\frac{e^{-i\mu l/2}}{\nu^{2}}\left(
\begin{array}{c}
(1-\mu^{2}) \left[ - 1 + i\mu z + \cosh \nu z
-\frac{i\mu}{\nu} \sinh \nu z \right] e^{i\mu x} \\ \\
\left[ \frac{\nu^2}{2} - i\mu(1-\mu^{2})z + \frac{\nu^2}{2} \cosh \nu z
+\frac{i\mu}{\nu} (3-7\mu^{2})\sinh \nu z \right] e^{-i\mu x}
\end{array}
\right) \nn\\ \nn\\
{\bf y}^{(3)}_{1} (x) &=&
\frac{e^{i\mu l/2}}{\nu^{2}}\left(
\begin{array}{c}
\left[ 2 i \mu + (1-\mu^{2})z - 2 i \mu \cosh \nu z
+\frac{1}{\nu} (1-5\mu^{2})\sinh \nu z \right] e^{i\mu x} \\ \\
(1-\mu^{2}) \left[ - z +\frac{1}{\nu} \sinh \nu z \right] e^{-i\mu x}
\end{array}
\right) \nn\\ \nn\\
{\bf y}^{(4)}_{1} (x) &=&
\frac{e^{-i\mu l/2}}{\nu^{2}}\left(
\begin{array}{c}
(1-\mu^{2}) \left[ - z +\frac{1}{\nu} \sinh \nu z \right] e^{i\mu x}
\\ \\
\left[ - 2 i \mu + (1-\mu^{2})z + 2 i \mu \cosh \nu z
+\frac{1}{\nu} (1-5\mu^{2})\sinh \nu z \right] e^{-i\mu x}
\end{array}
\right) ,
\nn
\eeq
where $\nu^{2} = 2(1-3\mu^{2})$ and $z = x +l/2$.

Due to the condition $Y_1 (-l/2) =I_4$, it is easy to see that the solution
with initial conditions (\ref{incon}) reads
\beq
{\bf y}_1 (x) = u_{1,0,1}(-l/2) {\bf y}^{(1)}_1 (x) + u_{1,0,2} (-l/2)
{\bf y}^{(2)}_1 (x) +v_{1,0,1} (-l/2) {\bf y}^{(3)}_1 (x) +
v_{1,0,2} (-l/2) {\bf y}^{(4)}_1 (x).
\nn
\eeq
Therefore by finding the relevant initial conditions from (\ref{incon}) in
the $k \to 0$ limit, we can determine the correct normalisation for
${\bf y}_{1} (x)$. The result is given by (\ref{zm_norm}).

\end{appendix}

\end{document}